\documentstyle[prl,aps,epsf]{revtex}
\newcommand{\upp}[1]{^{\rm \scriptscriptstyle #1}}
\newcommand{\dnn}[1]{_{\rm \scriptscriptstyle #1}}
\newcommand{\ket}[1]{|  #1 \rangle}
\newcommand{\bra}[1]{ \langle #1  |}
\newcommand{\proj}[1]{\ket{#1}\bra{#1}}

\newcommand{\tr}{{\rm Tr}}

\newcommand{\beq}{\begin{equation}}
\newcommand{\eeq}{\end{equation}}
\newcommand{\rank}[1]{{\cal R}(#1)}
\newcommand{\id}{\mbox{\boldmath{1} \hspace{-0.42cm} \boldmath{1}$\!\!$ }}
\newcommand{\be}{\begin{equation}}
\newcommand{\ee}{\end{equation}}

\newtheorem{theo}{Theorem}
\newtheorem{cor}{Corollary}
\newtheorem{conj}{Conjecture}

\title{Rank Two Bipartite Bound Entangled States Do Not Exist}
\author{Pawel Horodecki$^1$, John A. Smolin$^2$,
Barbara M. Terhal$^2$ and Ashish V. Thapliyal$^3$}
\address{\vspace*{1.2ex}
        \hspace*{0.5ex}{$^1$Faculty of Applied Physics and Mathematics,
Technical University of Gda\'nsk, 80-952 Gda\'nsk, Poland 
{\tt pawel@mifgate.pg.gda.pl}}}
\address{\vspace*{1.2ex}
        \hspace*{0.5ex}{$^2$IBM Research Division, Yorktown Heights, NY 10598,
        USA, {\tt smolin@watson.ibm.com, terhal@watson.ibm.com}}}
\address{\vspace*{1.2ex}
        \hspace*{0.5ex}{$^3$Dept. of Physics, Univ. of
California, Santa Barbara, CA 93106, USA, {\tt ash@physics.ucsb.edu}}}

\date{\today}
\begin{document}
\maketitle

\begin{abstract}
We explore the relation between the rank of a bipartite density matrix and the
existence of bound entanglement.  We show a relation between the rank,
marginal ranks, and distillability of a mixed state and use this to
prove that any rank $n$ bound entangled state must have support on no
more than an $n \times n$ Hilbert space. A direct consequence of this
result is that there are no bipartite bound entangled states of rank two.  
We also show that a
separability condition in terms of a quantum entropy 
inequality is associated with the above results.
We explore the idea of how many pure states are needed in a mixture to
cancel the distillable entanglement of a Schmidt rank $n$ pure state
and provide a lower bound of $n-1$.  We also prove that a mixture of a
non-zero amount of any pure entangled state with a pure product state
is distillable.
\end{abstract}


\medskip

Perhaps the central topic in quantum information theory has been the
study of entanglement, the nonclassical correlations between separated
parts of a quantum system.  In the early days of quantum theory
Einstein, Podolsky and Rosen discuss the paradoxical ``spooky action
at a distance'' of entangled particle pairs to express their disbelief
that quantum theory could provide a complete picture of reality \cite{EPR}.  Later,
Bell used entanglement to prove that quantum mechanics is inconsistent
with local reality \cite{bel:64}.
An understanding of entanglement seems to be at the heart of theories
of quantum computation and quantum cryptography \cite{BB84}, as it has been 
at the
heart of quantum mechanics itself.

In the case of bipartite pure states entanglement is rather well
understood, in the sense that there is a good measure of entanglement,
namely the entropy of the reduced density matrix of one party.
For a pure state $\ket{\psi}$ in a
Hilbert space belonging to two parties (traditionally named Alice and
Bob) ${\cal H}_A\otimes{\cal H}_B$ we define
\begin{equation}
E(\proj{\psi})=S(\tr_A \proj{\psi})=S(\tr_B \proj{\psi})
\label{epure}
\end{equation}
where $S$ is the von Neumann entropy function of a density matrix
given by $S(\rho)=-\tr (\rho \log \rho)$.  Different entangled bipartite
pure states of the same entanglement $E$ can be asymptotically
converted amongst one another while conserving entanglement
\cite{concentration}.  In the case of three or more parties sharing
entanglement, the situation is more complicated
(cf. \cite{ashish,multipartypaper}), as it is for mixed states even
for only two parties \cite{purification,BDSW}.

There are several measures of entanglement for bipartite mixed states.
The entanglement of formation $E_f$ is defined as
\begin{equation}
E_f(\rho)=\min \sum_{i=1}^k p_i E(\ket{\psi_i})
\label{eform}
\end{equation}
where the minimization is over all ensembles of pure states $\psi_i$ and
non-negative real numbers $p_i$ summing to 1 such that 
\begin{equation}
\sum_{i=1}^k p_i \proj{\psi_i}=\rho \ .
\end{equation}
The distillable entanglement $D(\rho)$ is defined as the asymptotic 
amount of pure
entanglement that can be gotten out of the mixed state $\rho$ using only
local quantum operations and classical communication (these were
defined originally in \cite{purification,BDSW}).  For pure states these
two measures are equal and equal to the entropy of the reduced density
matrix of each party as in Eq. (\ref{epure}).  For mixed states, we
have the property $D\le E_f$, reflecting the intuitive idea that one can't
distill more entanglement out of a state than was used in preparing
it.  Indeed, it has been shown \cite{horodeckibound} that there exist
states for which $D=0$ but $E_f>0$.  Such states are known as the {\em
bound entangled} states.

The entanglement of formation and the related relative entropy measure
of entanglement \cite{VPRK} share the problem that it is unknown
whether or not they are {\em additive}.  Is the entanglement of
formation of two copies of a state $\rho^{\otimes 2}=\rho \otimes
\rho$ always exactly twice that of one copy?  More generally, is the
entanglement of formation of $n$ copies of a state always exactly $n$
times the entanglement of formation of just one copy?  This has turned
out to be a rather difficult problem to solve, but a conjecture about
the bound entangled states of \cite{horodeckibound} may give us a
clue.  Bound entangled states have finite entanglement of formation,
but might it be that in the asymptotic limit as $n$ approaches
infinity we have $E_f(\rho_{\rm bound}^{\otimes n})/n\rightarrow 0$,
 which would explain why no entanglement can be distilled from them?
\footnote{Since the initial submission of this paper, Vidal and Cirac
have shown \cite{VCbound} that bound entangled states do exist that
have a nonzero asymptotic entanglement of formation (which is equal to
the entanglement cost \cite{EntCost}).  Still, the motivation remains to
look for low rank bound entangled states as a means of finding states
with subadditive entanglement of formation or even asymptotically zero
but single-copy finite entanglement of formation.}

One way of exploring this question is to numerically calculate the
entanglement of formation for a bound entangled state $\rho$, and then
for $\rho^{\otimes 2}$, $\rho^{\otimes 3}$, etc., and look for
subadditivity.  We have done this for some small examples, but
unfortunately the difficulty of the calculation scales exponentially
with the number of copies of $\rho$, and our only results so far have
been negative.  On the other hand, the rate of the exponential growth
is strongly dependent on the rank of the density matrix.

When trying to find the minimum ensemble (as in Eq. (\ref{eform}))
for a density matrix of rank
$R$ it is known that at most $k=R^2$ pure states $\psi_i$ are required
\cite{uhlmann} and each pure state is of dimension $R$.  Thus, for a
density matrix of the form $\rho^{\otimes n}$ with $\rho$ of rank $r$,
$\rho^{\otimes n}$ will have rank $R=r^n$ and the minimization will in
general require $r^{2n}$ vectors of dimension $r^n$.  For this reason
it is desirable to find bound entangled states of low rank.  The bound
entangled states of smallest known rank have rank four \cite{upb1}.

Recently results have been obtained suggesting 
the sensitivity of  bound entanglement to rank. 
In \cite{KZ} it was shown by means of numerical 
analysis that randomly generated bound entangled states 
in $2 \times 4$ typically have a participation 
ratio (a quantity related to the rank) $\tilde{R}\equiv 1/\tr(\varrho^2)$  
between 5 and 6. 
On the other hand for $2 \times n$ it was proved \cite{LCK} that all 
states which remain positive semi-definite under partial transposition
\footnote{The partial transpose is defined by
\begin{equation}
{\rm PT}(\rho)_{ij,kl}=\rho_{il,kj} 
\end{equation}
where the $i$ and $k$ indices are associated with Hilbert space ${\cal H}_A$ 
and the $j$ and $l$ indices with ${\cal H}_B$.
}
\cite{per:96} (PPT states) of rank $n$ and full rank of the reduced 
density matrix of the second party are separable.

In this paper we prove that if a bipartite density matrix's
rank is less than the rank of the marginal density matrix of either
party, then the density matrix has distillable entanglement (is not bound).
We use this to prove the negative result that there do not exist bound
entangled states of rank two, and to put some restrictions on what a rank
three bound entangled state must be like.  
Note that this implies that there are no unextendible product bases (UPBs)
\cite{upb1,upb2} in $m \times n$ with $m n -2$ members because the
complementary state to such a UPB would be rank 2 and bound entangled.

We also show that a mixture of a pure product state with a
non-vanishing amount of any pure entangled state is distillable and hence
not bound entangled. 
Finally we conjecture that any
irreducible bound entangled state in $n \times n$ has rank greater
than $n$ and show that this would imply that no rank three bound 
entangled state exists.

We will first prove a powerful theorem relating the distillability,
rank and marginal ranks of a mixed state.  Then we will use this
result to prove that there exists no bound entangled state of rank
two.

Consider a bipartite density matrix $\rho$ whose two parts belong to
Alice and to Bob.  We denote its {\em marginal or local density
matrices} by $\rho\upp{A}=\tr_B(\rho)$ and $\rho\upp{B}=\tr_A(\rho)$
respectively obtained by tracing out Bob and Alice. Its {\em marginal
rank} on Alice's side $\rank{\rho\upp{A}}$ is the rank of
$\rho\upp{A}$ and similarly for Bob. For pure states the marginal
ranks are equal and are also called the Schmidt rank of
the state \cite{asherbook}.  We say a state $\rho$ in $m \times n$ is
{\em irreducible} if and only if $\rank{\rho\upp{A}}=m$ and
$\rank{\rho\upp{B}}=n$. Intuitively this means that the density matrix
fully utilizes each of the local Hilbert spaces of Alice and Bob.

In our proof we use the reduction criterion of separability and
distillability \cite{hor:hor:97}:  If a state $\rho$
is separable then $\id \otimes \rho\upp{B} - \rho \ge 0$ and 
$\rho\upp{A} \otimes \id - \rho \ge 0$.  (A Hermitian matrix $H$
is positive semi-definite, or $H \ge 0$ for short, if and only if
$\bra{\psi}H\ket{\psi} \ge 0\ \forall_\psi$, or equivalently
$H$ has no negative eigenvalues.)  If this criterion is violated
then $\rho$ is distillable.  This provides a necessary condition for
separability and a sufficient condition for distillability.


\begin{theo} 
\label{theo:main}
If $\rank{\rho} < \max[\rank{\rho\upp{A}},\rank{\rho\upp{B}}]$ 
then $\rho$ is distillable. 
\end{theo}
{\bf Proof:} Without loss of generality let
$R=\rank{\rho\dnn{A}} > \rank{\rho}=r$. By local filtering
Alice takes $\rho$ to $\rho\dnn{f}$ such that,
\begin{equation}
\rho\dnn{f}\upp{A}= \frac{1}{R}\,\id\,_R \enspace.
\end{equation}
This can always be done with non-zero probability 
by applying the local filter 
$W=\sum_{i=1}^R \frac{1}{\mu_i R} \proj{\mu_i}$, where $\mu_i$ are the 
non-zero eigenvalues and $\ket{\mu_i}$ are the 
corresponding eigenvectors of $\rho\upp{A}$\cite{hor:hor:97}. 
Thus the eigenvalues of $\rho\dnn{f}\upp{A}$ are $1/R$. 
Since $\rho\dnn{f}$ has unit trace and is of rank $r$, its
largest eigenvalue $\lambda_{\rm max}$ cannot be less than $1/r$ which 
in turn is larger than $1/R$. 
Choosing $\ket{\psi}$ to be the eigenvector of $\rho\dnn{f}$ corresponding 
to the eigenvalue $\lambda_{\rm max}$ we have
\begin{equation}
\bra{\psi} \rho\dnn{f}\upp{A}\otimes \id - \rho\dnn{f} \ket{\psi} =
 \bra{\psi} \rho\dnn{f}\upp{A}\otimes \id \ket{\psi} - \lambda_{\rm max}  
\leq \frac{1}{R}-\frac{1}{r} < 0
\enspace.
\end{equation}
Thus the reduction criterion for separability  is violated and
hence $\rho\dnn{f}$ is distillable. Since $\rho\dnn{f}$ can be obtained 
from $\rho$ by a local quantum operation (filtering), this implies 
$\rho$ is distillable.  $\Box$

\noindent {\em Remark:}  It is also easy to show by the same
reasoning that if $\rank{\rho}=R$ then $\rho$ is distillable
except in the case where $\rho\dnn{f}$ is proportional to the
identity.  This is the case where $\lambda_{\rm max}$ is smallest: 
$\lambda_{\rm max}=1/R=1/r$.

Turning around the inequality (Theorem \ref{theo:main})
we have that if a mixed state $\rho$ is separable 
or bound entangled ({\em i.e.} not distillable) then 
\begin{equation}
\rank{\rho} \ge \max[\rank{\rho\upp{A}},\rank{\rho\upp{B}}].
\label{turnedaround}
\end{equation}
Using the monotonicity of logarithms, one can rephrase the theorem 
as an entropy inequality: 
For separable or bound entangled states
the entropy $S_0(\rho)\equiv \log(\rank{\rho})$ 
must satisfy the inequality     
\begin{equation}
S_{0}(\rho) \geq S_{0}(\rho\upp{A}){\rm\ and\ } 
S_{0}(\rho) \geq S_{0}(\rho\upp{B}) . 
\label{ei}
\end{equation}
The entropy $S_0$ (called the Hartley entropy) is a special case of quantum
Ren\'yi entropies
\begin{equation}
S_{\alpha}\equiv(1 -\alpha)^{-1}\log (\tr\rho^\alpha).
\end{equation}
The counterparts of the separability condition 
(\ref{ei}) have already been proved for the case of $\alpha=2$ and the cases 
where $\alpha$ limits to $1$ and $\infty$ (see \cite{alpha2}). 
It is interesting to note that the ratio of
violation of (\ref{ei}) in the limit  $\alpha\downarrow 1$ 
(the von Neumann entropy) gives in the case of $2 \times 2$ Werner states  
the yield of the hashing method of distillation of entanglement
\cite{BDSW}.
\medskip

Let us now look at some implications of this result for separable
and for bound entangled states.  

\begin{cor}
A rank $n$ separable or bound entangled state in a Hilbert space $\cal
H$ has support in at most an $n\times n$ subspace of $\cal H$.  Further,
there is no rank two bound entangled state.
\label{liveinspace}
\end{cor} 

\noindent{\bf Proof :} The first statement follows directly from
Eq. (\ref{turnedaround}).  The second 
statement is a consequence of the first and the fact that in $2 \times 2$ 
every entangled state is distillable \cite{hor:hor:hor:97}. $\Box$

\medskip

An open question is whether there exists a rank three bound entangled
state.  We can put some constraints on the form of such a state: If a
rank three bound entangled state exists, Corollary \ref{liveinspace}
shows it must have support on no more than a $3\times 3$ subspace. 
It must also be that such a state is irreducible because no bound
entanglement can exist in $2\times 2$ or $2\times 3$
\cite{hor:hor:hor:97}.  Then, as in the remark above, $\rho\dnn{f}$ must be
proportional to the identity on a three-dimensional subspace in
$3\times 3$.

Given the fact that a rank $n$ bound entangled state must live in a
$n\times n$ subspace, one may ask whether there are irreducible bound
entangled states of rank $n$ in $n \times n$.  Of course there is no
rank two bound entangled state in $2 \times 2$, and all the
known examples of bound entanglement are not of this type. The closest
one can get to this among known examples is the UPB state of rank four
in $3\times 3$ \cite{upb1}. Thus we conjecture:

\begin{conj}
There are  no
irreducible bound entangled states of rank $n$ in $n \times n$, 
i.e.\ any rank $n$ bound-entangled state can be expressed in a bipartite
space of dimension $n \times (n-1)$ or $(n-1) \times n$.
\end{conj}

If this conjecture holds then there are no rank three bound entangled states at
all.

Theorem \ref{theo:main} also has consequences for the ``cancellation
of distillable entanglement.''  Suppose one has a Schmidt rank $n$ pure state
$\ket{\Psi}=\sum_{i=1}^n \sqrt{\mu_i} \ket{\psi_i^A}\otimes \ket{\psi_i^B}$.  
How many other
arbitrary pure states $\ket{\phi_j}$ are needed in a mixture $\rho=p_0
\proj{\Psi} + \sum_{j=1}^k p_j \proj{\phi_j}$ before $\rho$ stops
being distillable?  We have the following corollary:

\begin{cor}
\label{cor:mixing}
If $\rho=p_0 \proj{\Psi} + \sum_{j=1}^{n-2} p_j \proj{\phi_j}$
with $p_0>0$ and $\ket{\Psi}$ of Schmidt rank $n$ then $\rho$ 
is distillable.
\end{cor}

\noindent{\bf Proof:} The proof follows from Theorem \ref{theo:main} and
the fact that $\rank{\rho}\le n-1$ and $\rank{\rho\upp{A}} \ge n$
since $\ket{\Psi}$ is Schmidt rank $n$.  Thus 
$\rank{\rho} \le n-1 < n \le \rank{\rho\upp{A}}$ and the theorem applies.
$\Box$

For $n=2$ the above corollary gives the empty result that a
Schmidt rank two state mixed with zero other states is distillable.
We will show a slight extension of this namely: 
\begin{theo}
\label{theo:mixing}
A mixture  of a non-zero amount of any entangled pure state with any
pure product state is always distillable. 
\end{theo}

\noindent{\bf Proof:} Consider the entangled and product pure states
in the Hilbert space of ${\cal H}_A \otimes {\cal H}_B$.  Since the
state $\rho$ made by mixing them has rank two, it is always
distillable unless $\rho$ has support on at most a $2\times 2$
subspace, by Corollary \ref{liveinspace}.  Thus, without loss of
generality we may choose the product state to be $\ket{0}\ket{0}$ and
the entangled state to be
$\ket{\psi}=a\ket{0}\ket{0}+b\ket{0}\ket{1}+c\ket{1}\ket{0}+d\ket{1}\ket{1}$.

In ${\cal H}_2 \otimes {\cal H}_2$ a density matrix $\rho$ is 
entangled \cite{per:96} and distillable \cite{hor:hor:hor:97} if and only if
the partial transpose of $\rho$ is not  positive semi-definite.

We consider the mixture
\be
\rho= p \proj{00} +  \proj{\psi}\ .  
\ee 
This mixture would be normalized by a denominator of $1+p$ but this
will not affect positivity and so we omit it.

The partial transpose ${\rm PT}(\rho)$ is 
\begin{equation}
\rho' = {\rm PT}(\rho)= 
\left( \begin{array}{cccc}
p+|a|^2&a^*b&ac^*&bc^*\\
ab^*&|b|^2&ad^*&bd^*\\
a^*c&a^*d&|c|^2&c^*d\\
b^*c&b^*d&cd^*&|d|^2\\
\end{array}\right)
\end{equation}
If a matrix has a negative determinant then the matrix is not positive
semi-definite.
Expanding the determinant of $\rho'$ using Cramer's rule on the top
row we can write
\be
\det (\rho')=p \det (C_{11})  + 
\det (\rho'-{p} \proj{00}) = p \det (C_{11})  + \det ({\rm PT}(\proj{\psi}))
\ee
where $C_{11}$ is the $3\times 3$ matrix formed by leaving out the first
row and first column of $\rho'$.  The second term is always negative
since $\ket{\psi}$ is entangled (this is easily seen by doing the 
partial transpose in the Schmidt basis and noting the fact that
the eigenvalues and hence the determinant of the partial transpose 
are invariant of which
basis the partial transposition is done.).  
Writing out the first term as
\be
-p |d|^2 (|a|^2|d|^2 - ab^*cd^*-a^*bc^*d + |b|^2|c|^2)
= -p |d|^2 |(ad-bc)|^2
\ee
we see that it is always less than or equal to zero.
 Thus the determinant of $\rho'$ is negative, implying $\rho'$ is not
positive semi-definite and $\rho$ is distillable.
$\Box$

{\em Remark:} In contrast to this theorem, a mixture of two entangled 
pure state {\em can} be separable,
for example, the equal mixture of $\frac{1}{\sqrt{2}}(\ket{00}+\ket{11})$
and $\frac{1}{\sqrt{2}}( \ket{00}-\ket{11})$.

We have explored the relation between the rank of a density
matrix and the existence of bound entanglement. In particular
we find that a density matrix with rank smaller than either
of the marginal ranks is distillable.  This gave us the result that
any rank $n$ bound entangled state must belong to a $n \times
n$ subspace.  
A direct consequence of this result is that there are no
bound entangled states of rank two. 

We also pointed out that there is a separability criterion
in terms of a quantum entropy inequality which is naturally associated with
our results.
Further, we have explored the idea of how many
pure states are needed to cancel the distillable entanglement of a Schmidt
rank $n$ pure state and we provided a lower bound of $n-1$.  We also showed
the related result that a mixture of a non-zero amount of 
any entangled pure state with a pure product state is distillable.
Thus mixing with a single pure product state cannot prevent the 
distillability of an entangled pure state.

It should be noted that our results on bound entanglement hold
even for bound entangled states whose partial transpositions are not
positive semi-definite, should such states exist (for evidence 
that they may, see 
\cite{npt1,npt2}).  This can be seen
because our proofs are based directly on distillability.

It is an open question whether bound entangled states of rank three exist.
Another related question is whether states with rank equal to
the marginal ranks can be bound entangled or whether the rank needs to be 
strictly greater.


The work of J.A. Smolin, B.M. Terhal and A.V. Thapliyal has been
supported in part by the Army Research Office under contract numbers
DAAG55-98-C-0041 and DAAG55-98-1-0366.  P. Horodecki is supported by
the Polish Committee for Scientific Research, contract No.~2 P03B 103
16 and would like to thank M. Horodecki for helpful 
discussion.
The authors would like to thank the organizers of the AQIP99 conference
for providing an environment for us to collaborate and for financial
support.

\end{document}